\documentclass[10pt,conference]{IEEEtran}
\IEEEoverridecommandlockouts

\usepackage{amsmath} 
\usepackage{graphicx}
\usepackage{textcomp}
\usepackage{xcolor}
\usepackage{siunitx}

\usepackage{commath}
\usepackage{mathtools}

\usepackage{array}
\usepackage{arydshln}
\renewcommand{\arraystretch}{1.2}
\usepackage{tabularx}
\usepackage{hhline}
\usepackage{multicol}
\usepackage{booktabs}
\usepackage{threeparttable}
\usepackage{rotating}
\usepackage{multirow}

\usepackage{enumitem} 

\definecolor{color1}{RGB}{142,202,230}
\definecolor{color2}{RGB}{33,158,188}
\definecolor{color3}{RGB}{2,48,71}
\definecolor{color4}{RGB}{255,183,3}
\definecolor{color5}{RGB}{251,133,0}
\usepackage{tikz}       
\usetikzlibrary{
  arrows,
  positioning,
  matrix,
  calc,
  decorations.pathreplacing,
  decorations.pathmorphing,
  decorations.markings,
  decorations.text,
  shapes,
  backgrounds,
  shadows,
  trees,
  fit,
  snakes,
  patterns,
  mindmap,
  intersections,
  calendar,
  plotmarks,
  spy}

\usepackage[bookmarks=false]{hyperref}
\usepackage[capitalize]{cleveref}

\DeclareMathOperator*{\argmin}{arg\,min}

\ifdefined\blind
\newcommand{\huawei}{[hidden]}
\else
\newcommand{\huawei}{Huawei}
\fi

\newcommand{\twocol}[1]{\multicolumn{2}{c}{#1}}
\newcommand{\bestval}[1]{\textbf{#1}}
\newcommand{\distribInfo}[1]{\multirow{3}{*}{#1}}
\newcommand{\lat}{\multicolumn{1}{c}{P99}}
\newcommand{\tput}{\multicolumn{1}{c}{TPS}}

\def\BibTeX{{\rm B\kern-.05em{\sc i\kern-.025em b}\kern-.08em
    T\kern-.1667em\lower.7ex\hbox{E}\kern-.125emX}}
\begin{document}

\title{Deep Recommender Models Inference:\\Automatic Asymmetric Data Flow Optimization}

\ifdefined\blind
\author{\IEEEauthorblockN{Hidden Authors}
\IEEEauthorblockA{\textit{Hidden Affiliation}}
}
\else
\author{\IEEEauthorblockN{Giuseppe Ruggeri}
\IEEEauthorblockA{\textit{Huawei Technologies} \\
Zurich, Switzerland \\
giuseppe.ruggeri1@huawei.com}
\and
\IEEEauthorblockN{Renzo Andri}
\IEEEauthorblockA{\textit{Huawei Technologies} \\
Zurich, Switzerland \\
renzo.andri@huawei.com}
\and
\IEEEauthorblockN{Daniele Jahier Pagliari}
\IEEEauthorblockA{\textit{Politecnico di Torino} \\
Torino, Italy \\
daniele.jahier@polito.it}
\and
\IEEEauthorblockN{Lukas Cavigelli}
\IEEEauthorblockA{\textit{Huawei Technologies} \\
Zurich, Switzerland \\
lukas.cavigelli@huawei.com}
}
\fi

\maketitle

\begin{abstract}
Deep Recommender Models (DLRMs) inference is a fundamental AI workload accounting for more than 79\% of the total AI workload in Meta’s data centers. DLRMs' performance bottleneck is found in the embedding layers, which perform many random memory accesses to retrieve small embedding vectors from tables of various sizes. We propose the design of tailored data flows to speedup embedding look-ups. Namely, we propose four strategies to look up an embedding table effectively on one core, and a framework to automatically map the tables asymmetrically to the multiple cores of a SoC. We assess the effectiveness of our method using the Huawei Ascend AI accelerators, comparing it with the default Ascend compiler, and we perform high-level comparisons with Nvidia A100. Results show a speed-up varying from 1.5x up to 6.5x for real workload distributions, and more than 20x for extremely unbalanced distributions. Furthermore, the method proves to be much more independent of the query distribution than the baseline.
\end{abstract}

\begin{IEEEkeywords}
Deep Recommender Models, Inference Optimization, Automatic tables sharding, Asymmetric Data Flow, Table lookups acceleration
\end{IEEEkeywords}

\section{Introduction}

Deep Learning-based Recommender Models (DLRMs) are pivotal in personalizing user experiences on major web services such as Amazon, Facebook, YouTube, and Netflix, significantly contributing to the business success of these platforms. For instance, in 2018, DLRMs were responsible for up to 35\% of Amazon's revenue and influenced 75\% of movies watched on Netflix~\cite{chui2018notes,gupta_architectural_2020}. The widespread adoption of DLRMs translates into a huge inference workload, representing, for instance, more than 79\% of the total AI workload in Meta's data centers \cite{gupta_architectural_2020}.
DLRMs utilize a combination of dense and extremely sparse categorical features for tasks like Click-through Rate Prediction (CTR).
Using neural networks, they learn intricate interactions between user- and item-dependent features, such as clicks on products or ratings on movies.

\textit{Embedding layers} play a crucial role in DLRMs. They map categorical features ($m$ categories) to dense, fixed-sized vectors of $E$ elements (commonly 16 to 128), using look-up tables (LUTs) of shape $m\times E$. This process creates a significant performance bottleneck due to the large number of random accesses to the LUTs, stored in DRAM or Flash memory~\cite{gupta_architectural_2020}. Furthermore, each table is accessed multiple times: for each query in the batch, and potentially multiple times per query (e.g., for retrieving the last five purchased items). 
In real-world deployment, DLRMs are subject to Service-Level Agreement (SLA) requirements, which often impose trade-offs between worst-case latency, average throughput, and the overall cost of the deployment system.

The solutions in the literature are many, tackling the problem from different angles.
Some works present system-level optimizations using CPUs, GPUs, FPGAs, or a cluster of them interconnected with a proper network topology \cite{jiang_fleetrec_2021}.
MicroRec \cite{jiang_microrec_2021} exploits FPGAs equipped with fast High-bandwidth Memory (HBM), redesigning the data structures used to memorize embedding tables to reduce the number of look-ups needed, such as storing the cartesian product of two tables.
FleetRec \cite{jiang_fleetrec_2021} proposes a strategy to build a heterogeneous cluster exploiting FPGAs with HBM for fast look-ups, CPUs with large DRAM capacity for a few large embedding tables, and GPUs exclusively for the computational part of the DLRM.

Other works, closer to ours, focus instead on optimizing the \textit{data flow} of embedding look-ups.
EVStore \cite{kurniawan_evstore_2023} proposes an ad-hoc cache replacement policy tailored for the characteristics of the embedding look-ups, demonstrating that given a single look-up per table, the overall performance is significantly degraded when even a single one does not hit the cache. 
Lastly, AutoShard \cite{zha_autoshard_2022} proposes to solve a table sharding problem by partitioning the LUTs across multiple nodes through a reinforcement learning algorithm driven by a deep neural network for estimating the cost of each table.

We propose innovative \textit{data flow strategies} for efficient table look-up and \textit{workload partitioning} approaches on a generic multicore architecture with minimal constraints, such as multicore AI accelerators and GPUs. 
Our experimental results with the Huawei Ascend 910 accelerator show promising outcomes, including up to 6.5x speed-up compared to the data flow obtained with the default Ascend compiler and improved performance trade-offs under various DLRM workloads and input query distributions. High-level estimations suggest that the developed strategy is likely to work well on GPUs, too.

\section{Embedding Lookup Data Flow Design}

\subsection{Target Platform}
\begin{figure}
    \centering
    \includegraphics[width=\linewidth]{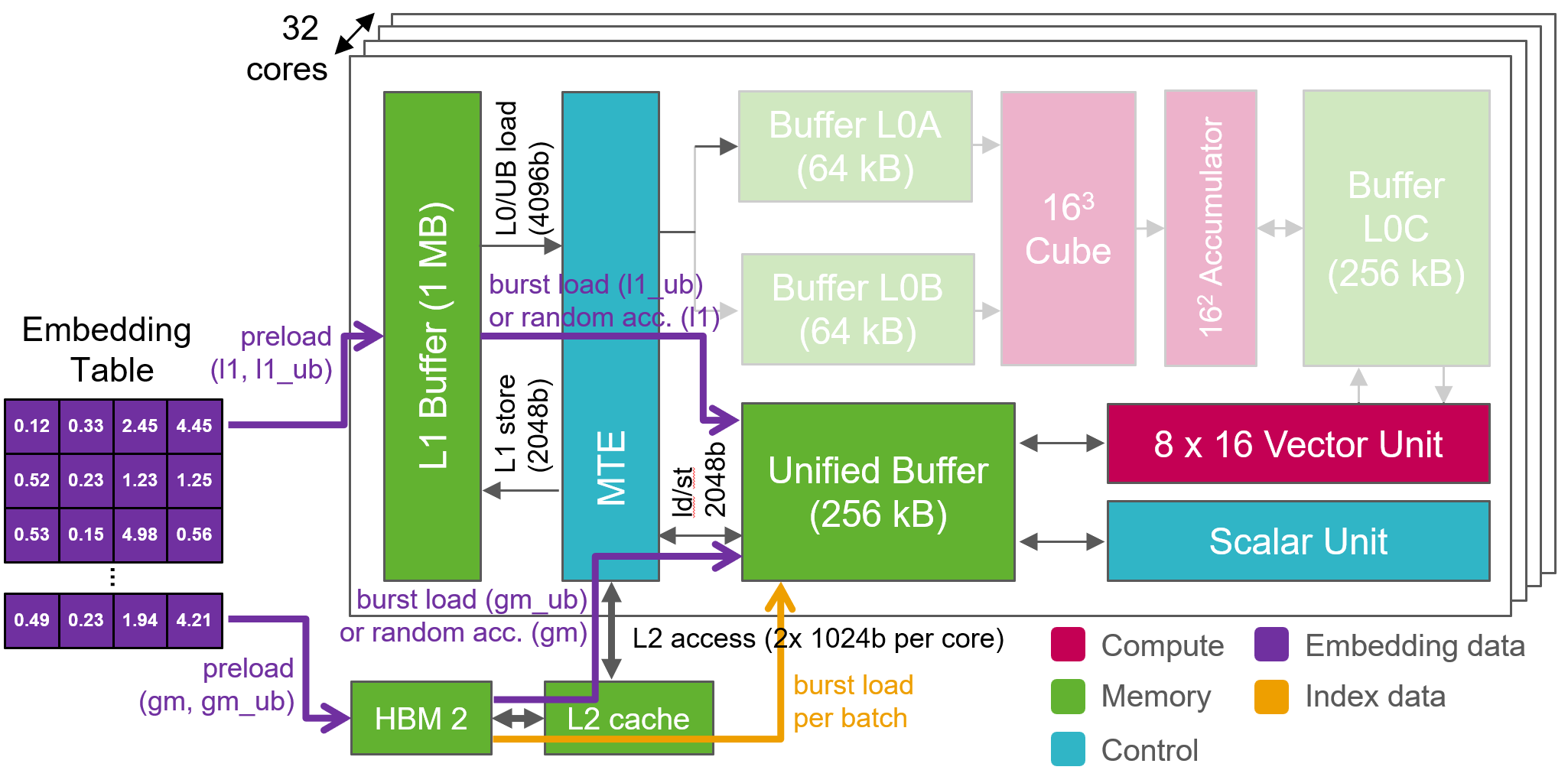}
    \caption{Key HW components and different data flows strategies on Ascend.}
    \label{fig:da_vinci_core}
\end{figure}
This work focuses on accelerating DLRM embedding look-ups on domain-specific multi-core hardware architectures for AI inference.
The architecture must meet a minimal set of requirements, namely it must include: 1) a shared memory to aggregate the looked-up rows; 2) a global memory (GM) to memorize the embedding tables, usually supported by a transparent layered cache; and 3) a vector compute unit that enables fast vectorized look-ups from shared-memory, as well as fast vectorized aggregations like vector sums.
Optionally, we propose to persist tables in shared memory when supported by the software stack.
AI accelerators and GPUs are representative examples of this type of platform. For our experiments, we target the Huawei Ascend AI accelerators, which are based on the DaVinci core, depicted in \cref{fig:da_vinci_core}. We also target the Nvidia A100 GPU based on the Nvidia Ampere architecture \cite{choquette_32_2021} to make high-level comparisons. 

\textit{Huawei Ascend Architecture}
As depicted in \cref{fig:da_vinci_core}, the Ascend architecture includes a scalar unit to execute arbitrary instructions, taking care of address computation and coordinating the other units.
The vector and cube units are instead designed to execute efficiently vector and matrix operations.
Importantly, there are two core-specific scratchpad memories, a 1\,MB L1 buffer, used as temporary fast storage close to the computational units, and a 256\,kB Unified Buffer (UB), positioned very close to the vector unit, all supported by Direct Memory Access (DMA) units.
Different SoC variants are built for different scenarios: embedded (Ascend 310), low-cost inference (Ascend 310P), and high-performance training \& inference (Ascend 910). Ascend 910 provides 32 cores sharing an on-chip L2 cache of 32\,MB, and has a 32\,GB HBM. A rich software stack is available, including a model compiler capable of converting high-level AI models into executable binaries applying layer fusion, data layout optimizations, and L2 cache locality optimizations~\cite{ascend_2021}. 

\subsection{Data Flow Design}
In DLRMs, inference performance is bottlenecked by the many random small memory accesses to the embedding layers' LUTs.
Specifically, for a sequence length of $s_i$ an embedding layer performs $s_i$ (sequence length) lookups from an embedding table of shape $(m_i,E)$ for each sample in a batch, with $m_i$ being the number of rows in the table, and $E$ the embedding dimension, thus retrieving $s_i$ vectors of $E$ elements. The sequence is then pooled together into a single vector of $E$ elements using a 
 reduction function (e.g., sum, average).
 
Given that: 1) HBMs waste a lot of bandwidth in retrieving many small vectors, 2) the bandwidth of the different memories in the hierarchy differs by orders of magnitude, and 3) the access patterns are extremely irregular~\cite{criteo-display-ad-challenge}, scratchpad memories should be exploited as much as possible to avoid data movements from off-chip memory. We propose four data flow strategies for accelerating look-ups and pooling operations based on two novelties: 1) preloading persistently the table in one of the memory buffers available (e.g., L1 buffer on Ascend) when supported by the software stack, and 2) vectorized look-ups using the vector unit, where the table is moved in chunks to the shared memory, and the vector unit retrieves multiple rows in parallel. Thus, not only do we reduce the bottleneck by accessing on-chip memory, but we also avoid the extreme dependence on the input query distribution using conflict-free memory accesses. The four strategies, depicted in \cref{fig:da_vinci_core}, are outlined below:

\noindent
\begin{itemize}[label={}, left=0pt, labelsep=0pt, itemsep=0pt, parsep=0pt]
    \item \textbf{GM and L1 strategy}. Read one row at a time (with double buffering) either from the off-chip memory (global memory - GM) or from the persistent buffer (L1) to the shared memory, followed by pooling this row in an accumulation buffer.
    \item \textbf{GM-UB and L1-UB strategy}. Performs the previously described vectorized look-up operations after moving the table in chunks to the shared memory. 
\end{itemize}

\section{Inter-core Workload Partitioning}

\subsection{Symmetric Partitioning}
The simplest inter-core partitioning scheme, referred to as symmetric partitioning, involves evenly splitting the workload across cores along the batch size dimension.
Given $N$ embedding tables, the table row counts $M = \{m_i\}$, the sequence lengths $S = \{s_i\}$, a batch size $B$, $K$ cores, the input query distribution $I$, and the persistent buffer size $L$, we define an objective function $J$ to minimize.
The problem becomes finding the optimal policy $\hat{P}=\{\hat{p}_0,\dots,\hat{p}_{N-1}\}$, where $\hat{p}_i\in\{\text{GM},\text{GM-UB},\text{L1},\text{L1-UB}\}$ is the data flow strategy applied for the $i$-th table, which minimizes the objective function $J$: 
\begin{align}
\label{eq:symm_problem_0}
    \hat{P} = \argmin_P\,\,J(B, M, S, K, L, I, P)
\end{align}
One option could be to define $J$ as the average latency. However, as we will discuss in \cref{sec:experiment_setup}, a better performance metric is the 99-th percentile (P99) of the acquired latencies.

The problem of defining $J$ is non-trivial as the execution time depends on many factors and is difficult to accurately model, as underlined in~\cite{zha_autoshard_2022}.
Thus, we assume that the execution of each embedding layer is independent of the other, and we estimate the P99 latency of each table with a simple linear formulation:
\begin{align}
\label{eq:symm_est}
    J_i = 
    \begin{cases}
        \beta_0 + \beta_1 \cdot \frac{B \cdot s_i}{K} &\text{if } p_i \neq \text{``UB''}\\
        \beta_0 + \beta_1 \cdot \frac{B \cdot s_i}{K} + \beta_2 \cdot m_i &\text{otherwise}\\
    \end{cases}.
\end{align}
The parameters $\beta$ in \eqref{eq:symm_est} differ for each configuration of the other (hyper-)parameters and are fitted using ordinary least squares on collected hardware measurements.

To solve the optimization problem in \eqref{eq:symm_problem_0} greedily, we first obtain each table's estimated costs for all four strategies based on the performance model. 
Then, we sort the tables by descending sequence length and ascending size, and we pick in order the tables that fit in L1 choosing the optimal strategy between L1 and L1-UB. Lastly, the optimal strategy between GM and GM-UB is chosen for the remaining tables. 

\subsection{Asymmetric Partitioning}
The symmetric partitioning is limited by having the same set of tables stored in the L1 of every core.
Accordingly, we propose to split the tables, and so the workload, in an asymmetric way.
This allows preloading different tables (or parts thereof) in the L1 buffer of different cores, thus achieving a usable aggregated L1 buffer $K$ times larger than the symmetric one.
The table chunks can be optionally replicated to split the batch workload. This is the first time such an approach has been used for DLRM inference.
For this purpose, we subtract the chunk's offset from the input indices and clip them to avoid out-of-bounds accesses.
Atomic inter-core accumulation is also needed since multiple cores handle the same batch portion, where an index may fall into different chunks.

\paragraph{Sharding and Load Balancing}
The many degrees of freedom of the asymmetric approach make policy optimization much more challenging.
Given a set of fixed costs, the table sharding problem is NP-hard~\cite{zha_autoshard_2022}, and consists in finding an optimal partitioning of the tables in $K$ groups to minimize the unbalancing between the total costs of the groups. Even worse, the number, size, and replication of chunks introduce additional complexity. 
Thus, we heuristically define a maximum chunk size and we split the tables into the least chunks while fixing the replication factor to 1.

Again, a greedy optimization algorithm guided by a linear performance model is proposed.
1) The tables larger than the L1 buffer are first split into the fewest chunks, but only if the speed-up offered by the L1 strategies exceeds the number of chunks;
2) They are then sorted by descending sequence length and ascending size, then allocated to the core with the lowest P99 execution time according to the performance model.
3) If there is L1 memory available, L1 or L1-UB are selected based on the model, otherwise, GM or GM-UB are chosen;
4) If the Load Imbalance Factor ($\text{LIF}=\frac{t_\text{max}}{t_\text{avg}}$) reaches a threshold, the remaining tables are partitioned symmetrically.

\section{Results}
\addtocounter{figure}{2}
\begin{figure*}
    \centering
    \includegraphics[width=0.422\linewidth]{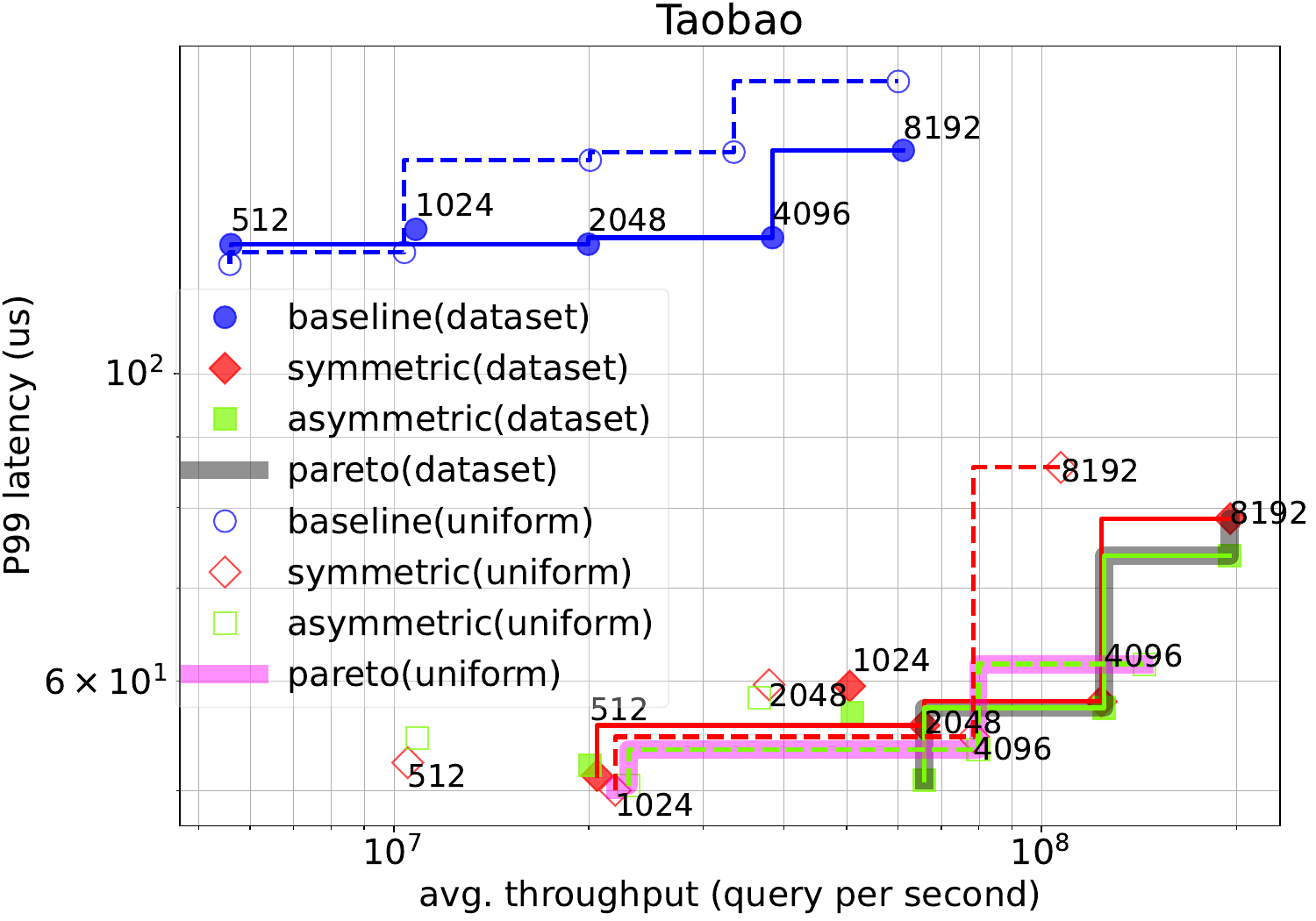}
    \includegraphics[width=0.422\linewidth]{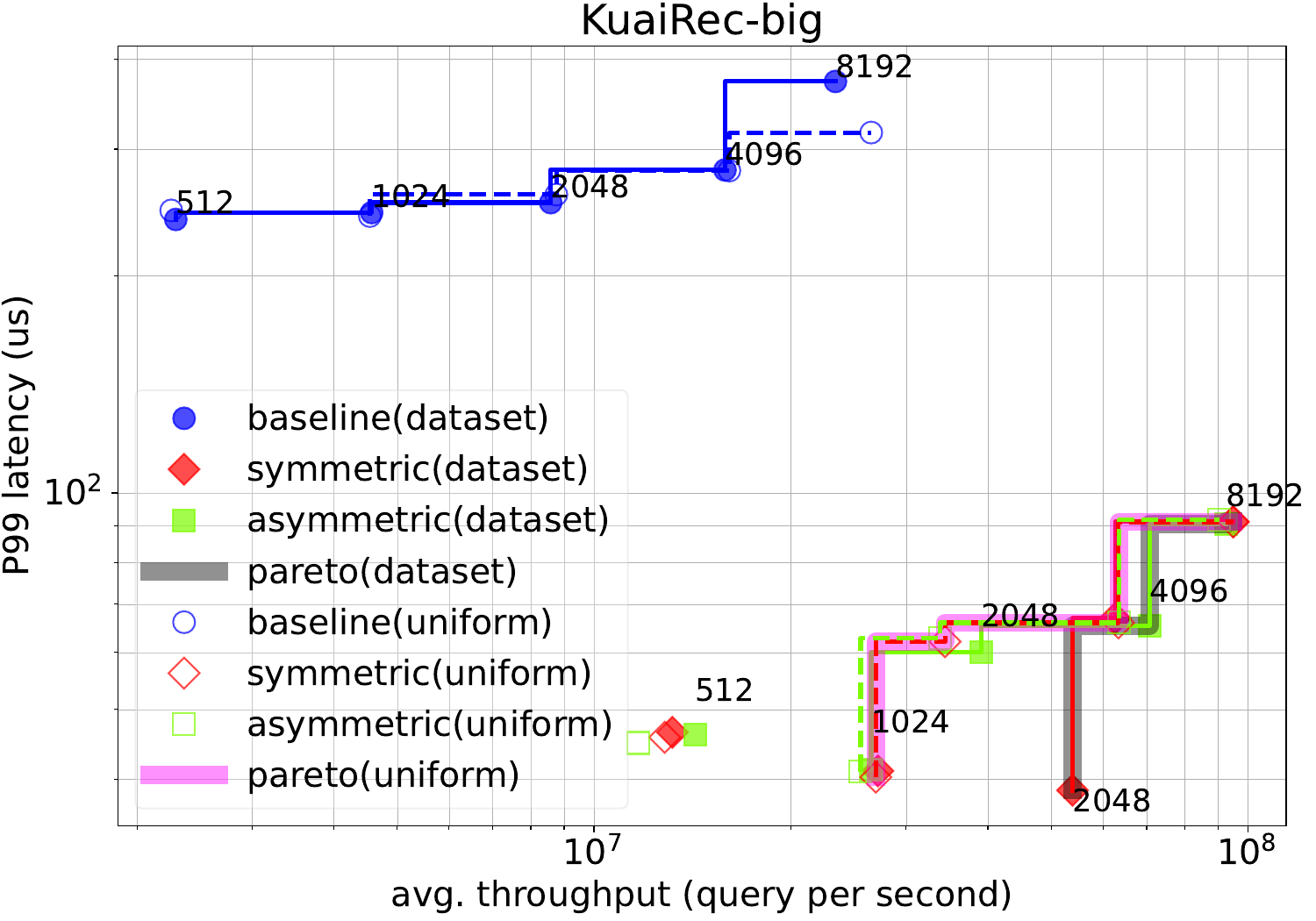}
    \caption{Avg. throughput [query/s] vs. P99 latency [\textmu s/batch] trade-offs varying the batch size on Ascend~910. The bold lines represent the global Pareto fronts.}
    \label{fig:perf_tradeoff}
\end{figure*}
\addtocounter{figure}{-3}

\subsection{Experimental Setup}
\label{sec:experiment_setup}
We profiled through hardware measurements various inference DLRM workloads using Ascend 910. Differently from the 310, this system is equipped with fast HBM, which can boost standard embedding lookups from off-chip memory, thus providing a stronger baseline.
We fixed the embedding dimension to 16 \texttt{float16} values and used the sum as reduction function.
We defined six workloads, each consisting of the extracted tables from the following datasets: Criteo-1TB~\cite{criteo-display-ad-challenge}, Avazu-CTR~\cite{avazu-ctr-prediction}, Taobao~\cite{chen2019serendipity}, and TenRec~\cite{yuan_tenrec_2022} for CTR tasks, and KuaiRec~\cite{gao_kuairec_2022} for video recommendations.
We further included a production DLRM by \huawei{}, referred to as \textit{\huawei{}-25MB}, for which no access distributions are available.

\ifdefined\blind
\begin{figure}
    \centering
    \includegraphics[width=0.77\linewidth,trim={0 0 2mm 0}]{fig/model_tables_distribution-hidden.pdf}
    \caption{Histogram of tables by row count for various workloads.}
    \label{fig:model_tables_distr}
\end{figure}
\else
\begin{figure}
    \centering
    \includegraphics[width=0.77\linewidth,trim={0 0 2mm 0}]{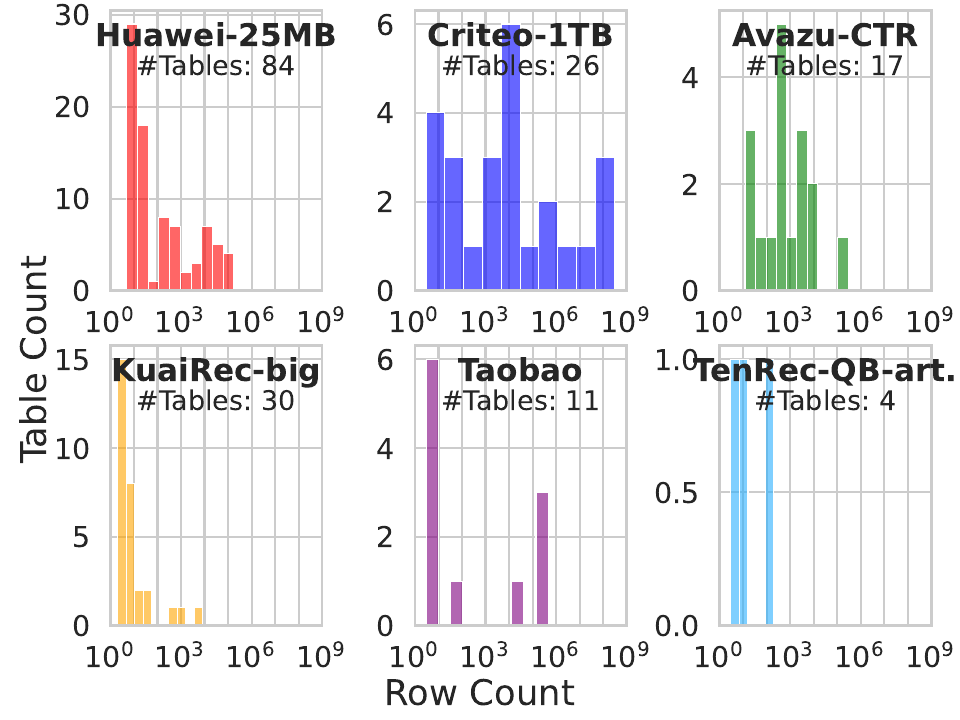}
    \caption{Histogram of tables by row count for various workloads.}
    \label{fig:model_tables_distr}
\end{figure}
\fi

\Cref{fig:model_tables_distr} shows that the table sizes of the workloads range from very small ones (\textless1\,MB), up to 26\,GB.
We excluded the \textit{user\_id} and \textit{item\_id} huge tables as we target only tables that fit in the global memory of the target hardware (32\,GB for Ascend910).
Our strategies could be extended to multiple devices for such tables, or alternatively, they could be looked up on the host side.
Additionally, the sequence length was fixed to 1 for all the workloads except \huawei{}-25MB, whose sequence lengths range from 1 up to 172.
We used three input query distributions: 1) \textbf{uniform random}, a stress test for caches, 2) \textbf{fixed}, all indices fixed to the same value, a stress test for bank conflicts, and 3) \textbf{pseudo-realistic}, randomly sampling according to the dataset's statistics.
Our baseline method is the vendor's official toolchain, generating a binary based on highly optimized built-in operators, complex layer fusion \& scheduling, and L2 cache locality optimizations.

\renewcommand{\arraystretch}{0.9} 
\setlength{\tabcolsep}{4pt} 
\begin{table*}[t]
\centering
\caption{P99 latency [s/batch] and average throughput (TPS) [query/s] with batch size 8192 on Ascend~910.}
\label{table:model_comparison}
\scriptsize
\begin{tabular}{ll rr rr rr rr rr rr} 
 \toprule
 \multirow{2}{1cm}{query\\distrib.} & \multirow{2}{*}{strategy} & \twocol{\huawei{}-25MB} & \twocol{Criteo-1TB} & \twocol{Avazu-CTR} & \twocol{KuaiRec-big} & \twocol{Taobao} & \twocol{TenRec-QB-art.} \\ 
  & & \lat & \tput & \lat & \tput & \lat & \tput & \lat & \tput & \lat & \tput & \lat & \tput \\ 
 \midrule
 \distribInfo{uniform} & baseline & 22872\textmu & 0.36M & 817\textmu & 15.8M & 223\textmu & 38M & 317\textmu & 26.8M & 163\textmu & 60M & 99\textmu & 87M\\
 & symmetric & 6020\textmu & 1.42M & \bestval{530\textmu} & \bestval{17.3M} & 69\textmu & 125M & \bestval{91\textmu} & \bestval{94.9M} & 86\textmu & 107M & 19\textmu & 501M\\
 & asymmetric & \bestval{5696\textmu} & \bestval{1.49M} & 583\textmu & 15.7M & \bestval{68\textmu} & \bestval{375M} & 92\textmu & 90.4M & \bestval{62\textmu} & \bestval{143M} & \bestval{17\textmu} & \bestval{512M}\\
 \midrule
 \distribInfo{real} & baseline & -- & -- & 1710\textmu & 4.89M & 765\textmu & 10.9M & 338\textmu & 24.9M & 145\textmu & 61M & 108\textmu & 71M\\
 & symmetric & -- & -- & 950\textmu & 9.9M & 406\textmu & 21.0M & 91\textmu & \bestval{94.9M} & 78\textmu & 195M & 19\textmu & 493M\\
 & asymmetric & -- & -- & \bestval{931\textmu} & \bestval{10.4M} & \bestval{333\textmu} & \bestval{24.6M} & \bestval{90\textmu} & 92.5M & \bestval{74\textmu} & \bestval{195M} & \bestval{17\textmu} & \bestval{496M}\\
 \midrule
 \distribInfo{fixed} & baseline & 155120\textmu & 104k & 538\textmu & 1.53M & 1314\textmu & 6.3M & 577\textmu & 14.4M & 1511\textmu & 5.71M & 375\textmu & 22M\\
 & symmetric & 55468\textmu & 177k & 2632\textmu & 3.43M & 445\textmu & 19.1M & \bestval{90\textmu} & \bestval{95.0M} & 982\textmu & 8.81M & 19\textmu & \bestval{497M}\\
 & asymmetric & \bestval{38203\textmu} & \bestval{216k} & \bestval{2148\textmu} & \bestval{3.98M} & \bestval{365\textmu} & \bestval{22.5M} & 93\textmu & 89.2M & \bestval{901\textmu} & \bestval{9.56M} & \bestval{18\textmu} & 492M\\
 \bottomrule
\end{tabular}
\end{table*}

DLRMs are usually deployed under SLA requirements, typically expressed as an upper bound on the 99th percentile of the latencies (P99 or worst-case latency) since rare failures do not have catastrophic consequences.

The study evaluates the trade-offs between average throughput (queries per second) and P99 latency, which is crucial as the DNN processing the embedding vectors is often adjusted to maximize prediction quality (i.e., revenue or user satisfaction) based on the remaining time available.

\subsection{High-level estimation comparison}

\ifdefined\blind
\newcommand{\highlevelsimpath}{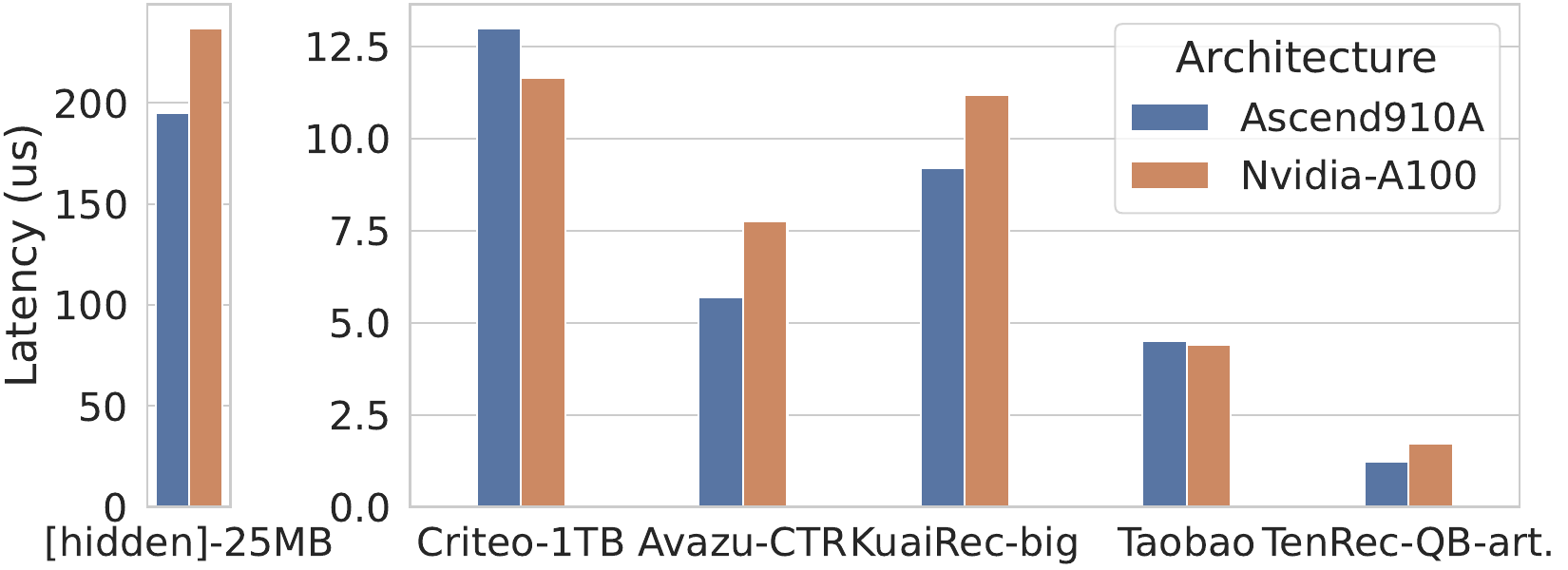}
\else
\newcommand{\highlevelsimpath}{fig/high-level-simulation-False.pdf}
\fi

\begin{figure}
    \centering
    \includegraphics[width=0.8\linewidth]{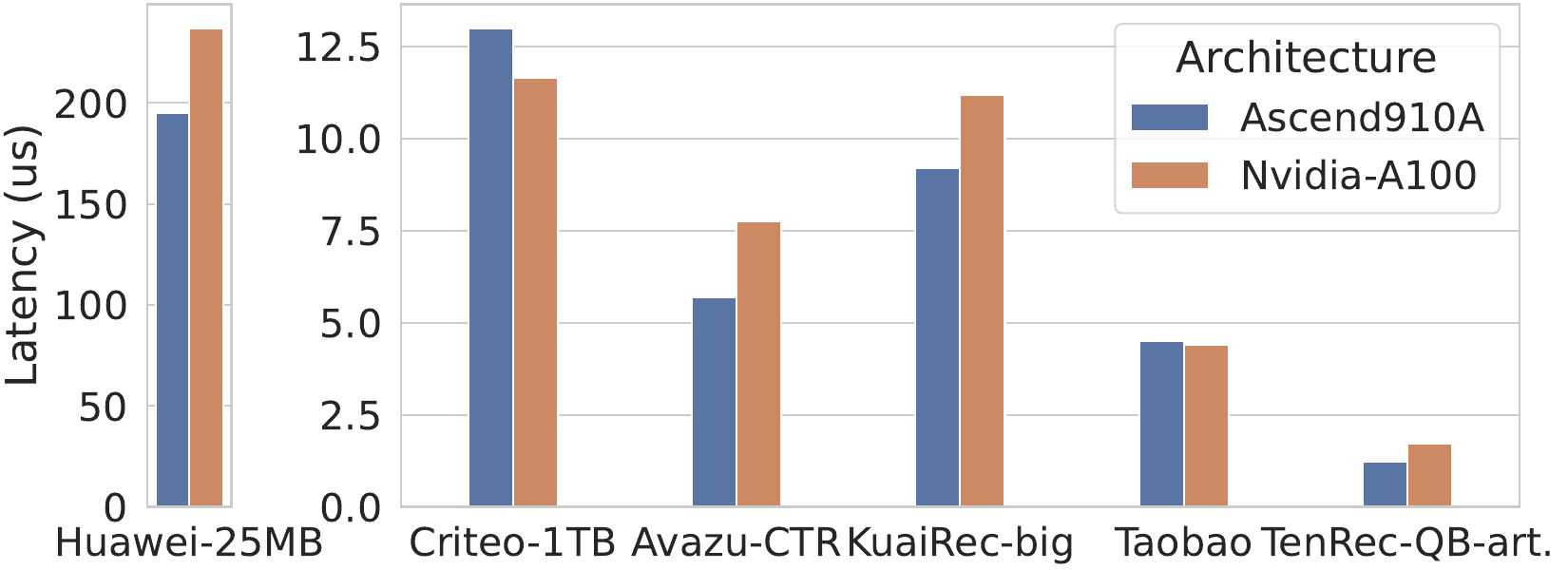}
    \caption{High-level performance estimation using a batch size of 8192. }
    \label{fig:high-level-model-comparison}
\end{figure}
To get an idea of the theoretical speed-up achievable with a given architecture, we compare the theoretical performance of Ascend~910 and Nvidia A100 based on the declared hardware specs.
The approximation assumes conflict-free memory accesses, symmetric partitioning, and no L1 persistent preloading for Nvidia A100 since it is not supported by the software stack.

\cref{fig:high-level-model-comparison} shows that both architectures deliver comparable performance for all workloads, thus suggesting that the proposed method could work well also on GPUs.
The estimation shows that Ascend delivers 1.2--1.3x better performance in most of the workloads, probably thanks to the L1 strategy exploitation. The limited performance difference is motivated by the tooideal hypothesis of conflict-free memories and the worse HBM of Ascend~910. In real executions, accesses may need to be sequentialized due to conflicts, and performing many small random memory accesses results in a much lower bandwidth than the maximum declared.

\subsection{Model-level Evaluation}
Comparing the trade-offs (P99 latency vs. average throughput) on Ascend~910 on two workloads and two query distributions, for varying batch sizes, reveals that our strategies provide consistently better trade-offs than the baseline, as shown in \Cref{fig:perf_tradeoff}. 
In particular, the asymmetric strategy provides the best performance for almost all the considered points.
Results with the dataset distribution are superior to the uniform one, probably due to the higher L2 cache hit ratio in the case of the GM and GM-UB strategies. The asymmetric strategy also provides more consistent performance across distributions, showing to be a much better candidate for real deployment scenarios.

\Cref{table:model_comparison} shows that our method achieves 1.5x to 6.5x lower P99 latency than the baseline on real distributions on all workloads, demonstrating the effectiveness in real use cases. Using a fixed distribution results in more than one order of magnitude latency increase using the baseline, due to the L2 cache conflicts of multiple cores trying to access the same cache line concurrently, causing the sequentialization of the accesses. Our method has much fewer conflicts thanks to L1 persistent preloading and burst transfers from global memory.
Overall, the proposed strategy achieves much better trade-offs than the baseline, thus translating into the possibility of having higher average throughput under a given P99 latency constraints and thus, increased revenue.

\section{Conclusion}
DLRMs directly impact ad revenue and customer satisfaction, but users' latency tolerance limits the usage of larger and more accurate models.
We show that efficiently gathering embedding vectors from smaller tables in on-chip buffers is essential, saving memory bandwidth for those tables that cannot fit on-chip.
The proposed method proves to be consistent across query distributions and reaches better performance trade-offs than the baseline on almost all workloads.

\bibliographystyle{IEEEtran}
\bibliography{IEEEabrv,./bibliography}

\end{document}